\documentclass[12pt]{article}
\usepackage{amssymb,amsmath,epsfig}
\usepackage{cite}
\allowdisplaybreaks

\begin{document}
\title{\bf Study of Complexity Factor and Stability of Dynamical Systems in $f(\mathcal{G})$ Gravity}

\author{M. Zeeshan Gul$^{1,2,3}$ \thanks{mzeeshangul.math@gmail.com},~ W. Ahmad$^{3}$ \thanks{waseemue.math1@gmail.com}
, M. M. M. Nasir$^{4}$
\thanks{muddassarnasir6666@gmail.com}, Faisal Javed$^{6}$
\thanks{faisaljaved.math@gmail.com},\\Orhan Donmez$^{5}$
\thanks{orhan.donmez@aum.edu.kw}
and Bander Almutairi$^{7}$ \thanks{balmutairi@ksu.edu.sa}\\
$^1$ College of Transportation, Tongji University, Shanghai 201804,
China.\\
$^2$ Research Center of Astrophysics and Cosmology, Khazar
University,\\ Baku, AZ1096, 41 Mehseti Street, Azerbaijan.\\
$^3$ Department of Mathematics and Statistics, The University of Lahore,\\
1-KM Defence Road Lahore-54000, Pakistan.\\
$^4$ Hajvery University Gulberg 3, Lahore, Pakistan.\\
$^5$ College of Engineering and Technology,
American University of the \\Middle East, Egaila 54200, Kuwait.\\
$^6$ Department of Physics, Zhejiang Normal University, Jinhua 321004,\\
People's Republic of China.\\
$^7$ Department of Mathematics, College of Science, King Saud
University, \\P.O. Box 2455, Riyadh 11451, Saudi Arabia.}

\date{}
\maketitle

\begin{abstract}

In this paper, we evaluate the complexity of the non static
cylindrical geometry with anisotropic matter configuration in the
framework of modified Gauss-Bonnet theory. In this perspective, we
calculate modified field equations, the $C$ energy formula and the
mass function that help to understand the astrophysical structures
in this modified gravity. Furthermore, we use the Weyl tensor and
obtain different structure scalars by orthogonally splitting the
Riemann tensor. One of these scalars, $Y_{TF}$ is referred to as the
complexity factor. This parameter measures the system's complexity
due to non-uniform energy density and non-isotropic pressure. We
select the identical complexity factor for the structure as used in
the non-static scenario, while considering the analogous criterion
for the most elementary pattern of development. This technique
involves formulating structural scalars that illustrate the
fundamental features of the system. A fluid distribution that
satisfies the vanishing complexity requirement and evolves
homologously is characterized as isotropic, geodesic, homogeneous,
and shear-free. In the dissipative scenario, the fluid remains
geodesic while exhibiting shear, resulting in an extensive array of
solutions.
\end{abstract}
\textbf{Keywords:} Stellar Structures; Complexity; Stability Analysis.\\
\textbf{PACS:} 04.40.-b; 04.40.Dg; 04.50.Kd

\section{Literature Review}

Einstein's theory of gravity transformed our understanding of
gravitational forces by providing explanations for cosmic events and
validating its precision through solar system studies. The existence
of singularities in general relativity ($\mathcal{GR}$) has
motivated scholars to investigate modifications to the
$\mathcal{GR}$. Modified theories of gravity are substitute models
that give profound truths and explanations for event such as dark
energy ($\mathcal{DE}$), accelerated expansion of the cosmos, and
other galactic as well as observational astronomy. Changing the
curvature invariants and the generic functions that go along with
them in the geometric section of the Einstein-Hilbert action is what
these theories are based on. These theories provide viable
explanations without an exotic form of energy. A wider variety of
dynamical behaviors, capable of simulating $\mathcal{DE}$ and
influencing the development of structures on a grand scale, can be
realized in this way. The $f(\mathcal{R})$ gravity is one of the
simplest forms of modified theories \cite{1}, the formulation of
this theory involves substituting the Ricci scalar with a generic
function from the Einstein-Hilbert action. A detailed analysis of
this theory and its consequences can be found in \cite{2}. There are
different forms of modified theories such as curvature, torsion and
non-metricity-based theories \cite{2a}-\cite{2ssss}.

One of the modified forms of $\mathcal{GR}$ is Lovelock gravity
which operates in a space of $n$ dimensions. It is found to be
comparable to $\mathcal{GR}$ in the case of four dimensions
\cite{3}. One Lovelock scalar is known as the Ricci scalar, and the
second is the Gauss-Bonnet ($\mathcal{GB}$) invariant. This
combination produces Einstein $\mathcal{GB}$ gravity in five
dimensions \cite{4}. Nojiri and Odintsov \cite{5} proposed the
concept of $f(\mathcal{G})$ gravity, which provides intriguing
characteristics of cosmic expansion. This theory is devoid of
instability issues such as phantom spin-2 instabilities \cite{6} and
shows consistency with cosmic structures \cite{7}. The exploration
of this theory and its influence on gravitational theories is a
lively and ongoing field of research. By including additional
$\mathcal{GB}$ terms, this theory delves into intriguing
cosmological effects that diverge from the expectations of
$\mathcal{GR}$. This theory aims to account for the observed cosmic
acceleration without $\mathcal{DE}$. Understanding the universe at
very small scales during the early cosmic era might benefit from
this modified proposal. By using 4D $\mathcal{GB}$ gravity, recent
theoretical and numerical studies have not only contributed to
understanding the observational results of various astrophysical
systems but also compared these findings with results obtained from
other gravity models \cite{7a}-\cite{7f}.

The behavior of stellar objects is significantly influenced by key
attributes of vast phenomena, such as matter, temperature, heat, and
other variables. Consequently, it is necessary to utilize a
mathematical formula that encompasses all important elements to
ascertain the complex nature of cosmic systems. The concept of
complexity, as outlined by complexity and data, was initially put
forward in \cite{8}. Initially, this concept was applied to flawless
crystals and perfect gases. In a perfect crystal, particles are
organized in a systematic arrangement, owing to their symmetrical
structure, resulting in minimal entropy. By contrast, the particles
in a perfect gas are scattered in a random manner, resulting in the
manifestation of maximum entropy. The existence of symmetry in a
perfect crystal implies that the probability distribution around its
symmetric component offers minimal information. A small fraction of
this probability is sufficient to describe all of its features.
Studying a small section of ideal gas allows for the acquisition of
the greatest quantity of information. Both of these structures
display distinct behavior when they are assigned an complexity of
zero. An additional concept of complexity was formulated by
examining deviations of different statistical states from the even
distribution of the structure that was considered \cite{9}.
Employing this notion, it is posited that ideal gases and perfect
crystals both show no complexity. In place of a probability
distribution, the new approach used energy density to assess the
complexity of astronomical objects \cite{10}. Nevertheless, this
benchmark did not integrate additional state parameters such as
heat, pressure, temperature, etc.

Herrera \cite{11} provided an updated definition of complexity in
relation to the  non-uniform energy density, pressure anisotropy and
Tolman mass is specific to static anisotropic fluid sources.
Different structural scalars are obtained by using orthogonal
decomposition on the Riemann tensor. The complexity factor is a
scalar data type that includes all the features described above.
Herrera and his colleagues \cite{13} generalized the notion of
complexity to reflect dissipative motion structure and examined two
forms of development. The complexity of the structure was also
calculated by the same authors with the assumption of axial symmetry
\cite{14}. Contreras et al \cite{15} analyzed the viability of the
developed models in both charged and uncharged cases by utilizing
the temporal aspect of the metric in Durgapal IV and V solutions,
under the constraint of vanishing complexity. A straightforward
method was developed by Contreras and Stuchlik in \cite{16} to
generate anisotropic interior solutions by the use of vanishing
conditions. The complexity of particular systems in non-minimally
$f(\mathcal{R})$ gravity was computed by Abbas and Nazar \cite{17}.
Using Herrera method, Nasir et al \cite{18} analyzed the impact of
dark source terms on the complexity of the static cylindrical
system.

In the presence of a strong gravitational field, evaluating the
solution requires examining the shift from spherical to asymmetrical
geometries. Levi Civita pioneered the concept of spacetime with a
cylindrical vacuum, which motivates astrophysicists to examine the
intriguing characteristics of different star systems. Herrera and
colleagues \cite{21} examined the cylindrical distribution by
analyzing scalar functions expressed as matter parameters. Adopting
modified $\mathcal{GB}$ gravity, Houndjo et al. \cite{22} developed
a set of seven solutions that correspond to three distinct feasible
models in cylindrical spacetime. Sharif and Butt \cite{23} employed
the Herrera approach to analyze the complex nature of cylindrical
spacetime by analyzing the Riemann tensor in the setting of static
matter configurations. Nasir and his collaborators discussed the
role of complexity for squared gravity \cite{25}.

The present work investigates the complexity of anisotropic
cylindrical spacetime in the framework of $f(\mathcal{G})$ theory.
The work is described as follows. In section \textbf{2}, we define
some basic definitions and compute non-zero components of the
modified field equations. Section \textbf{3} provides orthogonal
splitting of the Riemann tensor that yields four scalar functions.
Two evolutionary modes, namely homologous and homogeneous expansion
are discussed in section \textbf{4}. In section \textbf{5}, some
kinematical as well as dynamical quantities are derived to obtain
possible solutions in dissipative/non-dissipative modes. Section
\textbf{6} studies the stability of non-complex structures. We
summarize all our findings in section \textbf{7}.

\section{Gauss-Bonnet Theory}

This section addresses the physical parameters and modified field
equations to elucidate significant properties of self-gravitating
anisotropic fluid. The corresponding action of $f(\mathcal{G})$
theory is expressed as \cite{33}
\begin{equation}\label{1}
S=\int\bigg(\frac{\mathcal{R}+f(\mathcal{G})}{k}+L_{m}\bigg)\sqrt{-g}d^4x,
\end{equation}
where the Lagrangian density of matter distribution is defined by
$L_m$, the coupling constant is denoted by $k$ and the determinant
of the metric tensor is expressed by $g$. The resulting field
equations are
\begin{equation}\label{2}
\mathcal{R}_{\gamma\nu}+\frac{1}{2}g_{\gamma\nu}\mathcal{R}=8\pi{\mathbf{T}_{\gamma\nu}},
\end{equation}
where
\begin{equation}\label{3}
{\mathbf{T}_{\gamma\nu}}={\mathcal{T}^{(m)}_{\gamma\nu}}+\mathcal{T}^{(\mathcal{GB})}_{\gamma\nu}.
\end{equation}
In this relation, the matter component of $\mathcal{EMT}$ is denoted
by ${\mathcal{T}^{(m)}_{\gamma\nu}}$, while the modified component
arising from the effective terms of $\mathcal{GB}$ gravity is
defined by $\mathcal{T}^{(\mathcal{GB})}_{\gamma\nu}$, expressed as
\begin{eqnarray}\nonumber
\mathcal{T}^{(\mathcal{GB})}_{\gamma\nu}&=&\frac{1}{k}\bigg[(4\mathcal{R}_{\gamma\xi}
\mathcal{R}^{\xi} _{\nu}-2\mathcal{RR}_{\gamma\nu}
-2\mathcal{R}_{\gamma\xi\delta\eta}\mathcal{R}^{\xi\delta\eta}_{\nu}
+4\mathcal{R}_{\gamma\xi\nu\eta}\mathcal{R}^{\xi\eta})f_{\mathcal{G}}
\\\nonumber&+&
\frac{1}{2}{g_{\gamma\nu}}f(\mathcal{G})-2\mathcal{R}g_{\gamma\nu} \nabla^{2}
f_{\mathcal{G}}+2\mathcal{R}\nabla_{\gamma}\nabla_{\nu}f_{\mathcal{G}}
-4\mathcal{R}^{\xi}_{\gamma} \nabla_{\nu}\nabla_{\xi}f_{\mathcal{G}}
\\\nonumber&-&
4\mathcal{R}^{\xi}_{\nu}\nabla_{\gamma}\nabla_{\xi}f_{\mathcal{G}}
+4\mathcal{R}_{\gamma\nu}\nabla
^{2}f_{\mathcal{G}}+4g_{\gamma\nu}\mathcal{R}^{\xi\eta}\nabla_{\xi}
\nabla_{\eta}
f_{\mathcal{G}}\\\label{4}&-&4\mathcal{R}_{\gamma\xi\nu\eta}
\nabla^{\xi}\nabla ^{\eta}f_{\mathcal{G}}\bigg],
\end{eqnarray}
where $f_{\mathcal{G}}=\frac{df(\mathcal{G})}{d\mathcal{G}}$ and
${\nabla}^2$=${\nabla_\gamma}$${\nabla^\gamma}$ is the d'Alembert
operator. We assume an anisotropic matter configuration whose
energy-momentum tensor with heat flux $(q_{\gamma})$, four-velocity
$(\mathcal{V}_{\gamma})$ and four-vector $(\chi_{\gamma})$ is given
by
\begin{equation}\label{5}
{\mathcal{T}^{(m)}_{\gamma\nu}}=(\rho+\mathcal{P}_{\bot})
\mathcal{V}_{\gamma}\mathcal{V}_{\nu}+\mathcal{P}_{\bot}g_{\gamma\nu}
+(\mathcal{P}_{r}-\mathcal{P}_{\bot})\chi_{\gamma}\chi_{\nu}+q_{\gamma}\mathcal{V}_{\nu}
+\mathcal{V}_{\gamma}q_{\nu},
\end{equation}
The four-vector is introduced to describe the preferred spatial
direction in an anisotropic fluid. In an anisotropic fluid, the
pressure is not same in all directions, meaning that the
stress-energy tensor would not only involve the metric tensor and
the four-velocity. But, an additional spatial direction must be
specified to characterize the anisotropy. This is done using the
four-vector, which satisfies the normalization condition
$(\chi^{\gamma} \chi_{\gamma}=1$, ensuring that $\chi_{\gamma}$ is a
unit spacelike vector) and orthogonality with the fluid's
four-velocity $(\mathcal{V}^{\gamma} \chi_{\gamma}=0)$ which
determines that $\chi_{\gamma}$ demonstrates a spatial direction
relative to the observer comoving with the fluid. The term
$(\mathcal{P}_r-\mathcal{P}_{\bot})\chi_{\gamma} \chi_{\nu}$ in
Eq.(5) accounts for the pressure anisotropy in the fluid. If the
fluid is isotropic then $\mathcal{P}_r=\mathcal{P}_{\bot}$ and this
term vanishes. If the fluid is anisotropic then the pressure differs
along the direction of $\chi_{\gamma}$ and this term contributes to
the energy-momentum tensor.

We consider non-static cylindrical spacetime as \cite{33a}
\begin{equation}\label{6}
ds^{2}=-\mathcal{J}^{2}(t,r)dt^{2}+\mathcal{K}(t,r)^{2}dr^{2}+\mathcal{L}(t,r)^{2}(d\theta^{2}
+\alpha^2dz^{2}),
\end{equation}
where the term $\alpha$ appears in the metric component associated
with the $z$-coordinate is a constant quantity having dimension of
inverse length. Using Eqs.\eqref{2}-\eqref{6}, we obtain the field
equations as
\begin{align}\label{7}
8{\pi}\bigg(\rho+\frac{\mathcal{T}^{(\mathcal{GB})}_{00}}{\mathcal{J}^2}\bigg)&=\bigg(2\frac{
\dot{\mathcal{K}}}{\mathcal{K}}+\frac{\dot{\mathcal{L}}}{\mathcal{L}}\bigg)\frac{\dot{\mathcal
{L}}}{\mathcal{J}^2\mathcal{L}}-\frac{1}{\mathcal{K}^2}\bigg[2\frac{\mathcal{L}''}{\mathcal{L}}
+\bigg(\frac{\mathcal{L}'}{\mathcal{L}}\bigg)^{2}-2\frac{\mathcal{K}'\mathcal{L}'}{\mathcal{K}\mathcal{L}}\bigg],
\\\label{8}
4{\pi}\bigg(q-\frac{\mathcal{T}^{(\mathcal{GB})}_{01}}{\mathcal{J}\mathcal{K}}\bigg)&=
\frac{1}{\mathcal{J}\mathcal{K}}\bigg(\frac{\dot
{\mathcal{L}'}}{\mathcal{L}}-\frac{\mathcal{L}'\dot{k}}{\mathcal{K}\mathcal{L}}-\frac{\dot{\mathcal{L}}\mathcal{J}'}
{\mathcal{L}\mathcal{J}}\bigg),
\\\label{9}
8{\pi}(\mathcal{P}_{r}+\frac{\mathcal{T}^{(\mathcal{GB})}_{11}}{\mathcal{K}^{2}})&=\bigg(\frac{2
\mathcal{J}'}{\mathcal{J}}+\frac{\mathcal{L}'}{\mathcal{L}}\bigg)\frac{\mathcal{L}'}{\mathcal{K}
^2\mathcal{L}}-\frac{1}{{\mathcal{J}}^2}\bigg[2\frac{\ddot{\mathcal{L}}}{\mathcal{L}}-\bigg(2
\frac{\dot{\mathcal{J}}}{\mathcal{J}}-\frac{\dot{\mathcal{L}}}{\mathcal{L}}\bigg)\frac{\dot{
\mathcal{L}}}{\mathcal{L}}\bigg]
\\\nonumber
8{\pi}(\mathcal{P}_{\bot}+\frac{\mathcal{T}^{(\mathcal{GB})}_{22}}{\mathcal{L}^{2}})&=\frac{1}
{\mathcal{K}^{2}}\bigg[\frac{\mathcal{J}''}{\mathcal{J}}+\frac{\mathcal{L}''}{\mathcal{L}}-\frac
{\mathcal{J}'\mathcal{K}'}{\mathcal{J}\mathcal{K}}+\bigg(\frac{\mathcal{J}'}{\mathcal{J}}-\frac
{\mathcal{K}'}{\mathcal{K}}\bigg)\frac{\mathcal{L}'}{\mathcal{L}}\bigg],
\\\label{10}&-
\frac{1}{\mathcal{J}^{2}}\bigg[\frac{\ddot{\mathcal{K}}}{\mathcal{K}}+\frac{\ddot{\mathcal{L}}}
{\mathcal{L}}-\frac{\dot{\mathcal{J}}}{\mathcal{J}}(\frac{\dot{\mathcal{K}}}{\mathcal{K}}+\frac
{\dot{\mathcal{L}}}{\mathcal{L}})+\frac{\dot{\mathcal{K}}\dot{\mathcal{L}}}{\mathcal{K}\mathcal{L}}\bigg].
\end{align}
In these equations, the dot and prime denote the derivatives with
respect to temporal and radial coordinates, respectively. Now, we
define the shear tensor as
\begin{equation}\label{11}
\sigma_{\gamma\nu}=\mathcal{V}_{(\gamma;\nu)}+a_{(\gamma}\mathcal{V}_{\nu)}
-\frac{1}{3}{\Theta} h_{\gamma\nu},
\end{equation}
where acceleration is denoted by
$a_{\gamma}=\mathcal{V}^{\nu}\mathcal{V}_{\gamma;\nu}$ and expansion
scalar is represented by $\Theta=\mathcal{V}^{\gamma}_{;\gamma}$.
Using Eqs.\eqref{6} and \eqref{11}, we have
\begin{equation}\label{12}
a_{1}=\frac{\mathcal{J}'}{\mathcal{J}}, \quad
a=\sqrt{a^{\gamma}a_{\gamma}}=\frac{\mathcal{J}'}{\mathcal{J}\mathcal{K}},
\quad\Theta=\frac{1}{\mathcal{J}}\bigg(\frac{\dot{\mathcal{K}}}{\mathcal{K}}
+2\frac{\dot{\mathcal{L}}}{\mathcal{L}}\bigg),
\end{equation}
where the term $a_1$ manifests the radial component of the
acceleration vector in the cylindrical coordinate system. This
describes how the proper time measured by comoving observers varies
with radial position, contributing to the four-acceleration. The
nonzero components of shear tensor are
\begin{equation}\nonumber
\sigma_{11}=\frac{2}{3}\mathcal{K}^{2}\sigma, \quad \sigma_{22}=\frac{\sigma_{33}}{\sin^{2}\theta}
=-\frac{1}{3}\mathcal{L}^{2}\sigma,
\end{equation}
and the shear scalar is given by
\begin{equation}\label{14}
\sigma^{\gamma\nu}\sigma_{\gamma\nu}=\frac{2}{3}\sigma^{2},
\end{equation}
with
\begin{equation}\label{15}
\sigma=\frac{1}{\mathcal{J}}\bigg(\frac{\dot{\mathcal{K}}}{\mathcal{K}}
-\frac{\dot{\mathcal{L}}}{\mathcal{L}}\bigg).
\end{equation}
Equation (\ref{8}) is reformulated as follows by employing Eqs.(\ref{12}) and (\ref{15})
\begin{equation}\label{16}
4\pi\bigg(q-\frac{\mathcal{T}^{(\mathcal{GB})}_{01}}{\mathcal{J}\mathcal{K}}\bigg)
+\frac{\sigma\mathcal{L}'}{\mathcal{K}\mathcal{L}}=\frac{1}{3\mathcal{K}}(\Theta-\sigma)'.
\end{equation}

The C-energy (which can be interlinked with the mass function)
provides the mass of the cylindrical system as \cite{33b}
\begin{eqnarray}\label{16a}
\mathcal{E}=m(t,r)=\frac{1}{8}
\left(1-\ell^{-2}\nabla^{\gamma}r\nabla_{\gamma}r\right),
\end{eqnarray}
where $\ell^{2}= \varsigma_{(3)\gamma}\varsigma^{\gamma}_{(3)}$,
$r=\upsilon\ell$ and  $\upsilon^{2}=
\varsigma_{(2)\gamma}\varsigma^{\gamma}_{(2)}$ define the specific
length, circumference radius and areal radius of cylindrical
geometry. The entities
$\varsigma_{(2)}=\frac{\partial}{\partial\phi}$ and
$\varsigma_{(3)}=\frac{\partial}{\partial z}$ define the Killing
vectors. It serves as an analogue to the energy concept in Newtonian
mechanics but is adapted to the relativistic setting where
gravitational waves carry energy. The $C$-energy is also called the
cylindrical energy, introduced by Geroch in the study of
gravitational waves propagating along cylindrical symmetry. Unlike
the standard energy-momentum tensor description does not always
provide a well-defined local energy density for the gravitational
field, the $C$-energy offers a way to quantify the energy content of
cylindrically symmetric spacetimes. The $C$-energy provides a
measure of the energy content of cylindrical gravitational waves. It
can be used to study energy conservation and flux in radiative
cylindrical spacetimes. It plays a role in analyzing the nonlinear
evolution of gravitational waves in spacetimes with cylindrical
symmetry. Manipulating Eq.(\ref{16a}), we obtain
\begin{equation}\label{17}
m=\frac{\mathcal{L}}{2}\bigg[\frac{1}{4}+\bigg(\frac{\dot{\mathcal{L}}}
{\mathcal{J}}\bigg)^{2}-\bigg(\frac{\mathcal{L}'}{\mathcal{K}}\bigg)^{2}\bigg].
\end{equation}
The appropriate time derivative $(D_{\tau})$ is defined as
\begin{equation}\label{18}
D_{\tau}=\frac{1}{\mathcal{J}}\frac{\partial}{\partial{t}}, \quad
\mathcal{U}=D_{\tau}\mathcal{L}<0,
\end{equation}
where $\mathcal{U}$ denotes the velocity of the fluid which
corresponds to the derivative of the radius with respect to proper
time. By using $\mathcal{U}$, Eq.(\ref{17}) can be represented as
\begin{equation}\label{19}
E=\frac{\mathcal{L}'}{\mathcal{K}}=\bigg(\mathcal{U}^{2}+\frac{1}{4}-\frac{2m}
{\mathcal{L}}\bigg)^{1/2}.
\end{equation}
Substituting the value of $E$ into Eq.(\ref{8}), we have
\begin{equation}\label{20}
4\pi\bigg(q-\frac{\mathcal{T}^{(\mathcal{GB})}_{01}}{\mathcal{J}\mathcal{K}}\bigg)
=E\bigg[\frac{1}{3}D_{\mathcal{L}}(\Theta-\sigma)
-\frac{\sigma}{\mathcal{L}}\bigg],
\end{equation}
where
$D_{\mathcal{L}}=\frac{1}{\mathcal{L}'}\frac{\partial}{\partial{r}}$
defines the proper radial derivative. By using preceding equations
and taking proper time and proper radial derivatives of
Eqs.(\ref{17}), we have
\begin{align}\label{21}
D_{\tau}m&=-4\pi\bigg[\bigg(\mathcal{P}_{r}+\frac{\mathcal{T}^{(\mathcal{GB})}_{11}}{\mathcal{K}^2}
\bigg)\mathcal{U}+\bigg(q-\frac{\mathcal{T}^{(\mathcal{GB})}_{01}}{\mathcal{J}\mathcal{K}}\bigg)E\bigg]\mathcal{L}^{2}+\frac
{\dot{\mathcal{L}}}{8\mathcal{J}},
\\\label{22}
D_{\mathcal{L}}m&=4\pi\bigg[\bigg(\rho+\frac{\mathcal{T}^{(\mathcal{GB})}_{00}}{\mathcal{J}^2}
\bigg)+\frac{\mathcal{U}}{E}\bigg(q-\frac{\mathcal{T}^{(\mathcal{GB})}_{01}}{\mathcal{J}\mathcal{K}}\bigg)+\frac{1}{32\pi
\mathcal{L}^2} \bigg]\mathcal{L}^{2}.
\end{align}
By taking integration of Eq.(\ref{22}), expression for the mass function appears as
\begin{equation}\label{23}
\frac{3m}{\mathcal{L}^{3}}=4\pi\rho-\frac{4\pi}{\mathcal{L}^{3}}
\int^{r}_{0}\mathcal{L}^{3}
\bigg[D_{\mathcal{L}}(\rho)-\frac{3}{\mathcal{L}}\bigg\{\frac{\mathcal{T}^{(\mathcal{GB})}_{00}}
{\mathcal{J}^2}+
\frac{\mathcal{U}}{E}\bigg(q-\frac{\mathcal{T}^{(\mathcal{GB})}_{01}}{\mathcal{J}\mathcal{K}}\bigg)
\bigg\} \bigg]\mathcal{L}^{'}dr+\frac{3}{8\mathcal{L}^2}.
\end{equation}

\subsection{Analysis of Structure Scalars}

This section examines the concept of structure scalars in modified
$\mathcal{GB}$ gravity, which are essential to understand the
complexity factor. The structure scalars in $\mathcal{GR}$ were
formulated by Herrera et at in \cite{36}. The magnetic component of
the Weyl tensor vanishes, thus, we consider the electric component
as
\begin{equation}\label{24}
\mathbb{E}_{\gamma\nu} = \mathcal{C}_{\gamma\lambda\nu\xi}
\mathcal{V}^{\lambda}\mathcal{V}^{\xi}, \quad \lambda, \xi=0,1,2,3,
\end{equation}
where
\begin{equation}\label{25}
\mathcal{C}_{\gamma\nu\lambda\xi}=(g_{\gamma\nu\mu\sigma}g_{\lambda\xi\tau\chi}-\eta_{\gamma
\nu\mu\sigma}\eta_{\lambda\xi\tau\chi})\mathcal{V}^{\mu}\mathcal{V}^{\tau}\mathbb{E}^{\sigma\chi},
\quad\mu,\sigma,\tau,\chi=0,1,2,3,
\end{equation}
where $\eta_{\gamma\nu\mu\sigma}$ is Levi-Civita tensor and
$g_{\gamma\nu\mu\sigma}
=g_{\gamma\mu}g_{\nu\sigma}-g_{\gamma\sigma}g_{\nu\mu}$. Solving
Eq.\eqref{24}, we have
\begin{equation}\label{26}
\mathbb{E}_{11}=\frac{2}{3}\mathcal{K}^{2}\xi, \quad
\mathbb{E}_{22}=-\frac{1}{3}\mathcal{L}^{2}\xi, \quad
\mathbb{E}_{33}=\mathbb{E}_{22}\sin^{2}\theta,
\end{equation}
Also,
\begin{equation}\label{27}
\mathbb{E}^{\gamma\nu}=\xi\left(\frac{1}{3}h^{\gamma\nu}
+\chi^{\gamma}\chi^{\nu} \right).
\end{equation}
The term $h^{\gamma\nu}$ in Eq.\eqref{27} defines the projection
tensor or the induced metric on the hypersurface orthogonal to the
fluid's four-velocity. The projection tensor is crucial because it
projects any tensorial quantity onto the three-dimensional spatial
hypersurface that is locally orthogonal to the fluid's
four-velocity. The presence of projection tensor ensures that the
electric part of the Weyl tensor is decomposed into its spatially
projected components and its contribution along the preferred radial
direction

The value of $\xi$ in Eq.\eqref{27} is given by
\begin{eqnarray}\nonumber
\xi&=&\frac{1}{2\mathcal{J}^{2}}\bigg[\frac{\ddot{\mathcal{L}}}{\mathcal{L}}-\frac
{\ddot{\mathcal{K}}}{\mathcal{K}}-\bigg(\frac{\dot{\mathcal{L}}}{\mathcal{L}}-\frac{\dot
{\mathcal{K}}}{\mathcal{K}}\bigg)\bigg(\frac{\dot{\mathcal{J}}}{\mathcal{J}}+\frac{\dot
{\mathcal{L}}}{\mathcal{L}}\bigg)\bigg]-\frac{1}{2\mathcal{L}^{2}}\\\label{28}
&+&\frac{1}{2\mathcal{K}^{2}}\bigg[\frac{\mathcal{J}''}{\mathcal{J}}-\frac{\mathcal{L}''}
{\mathcal{L}}+\bigg(\frac{\mathcal{K}'}{\mathcal{K}}+\frac{\mathcal{L}'}{\mathcal{L}}\bigg)
\bigg(\frac{\mathcal{L}'}{\mathcal{L}}-\frac{\mathcal{J}'}{\mathcal{J}}\bigg)\bigg].
\end{eqnarray}
The scalar functions $\mathcal{X}_{TF}$ and $\mathcal{Y}_{TF}$
originate from the orthogonal decomposition of the Riemann tensor,
we define the tensor as
\begin{eqnarray}\label{29}
\mathcal{Y}_{\gamma\nu}&=&\mathcal{R}_{\gamma\lambda\nu\xi}\mathcal{V}^{\lambda}\mathcal{V}^{\xi},
\\\label{30}
\mathcal{Z}_{\gamma\nu}&=&_{*}\mathcal{R}_{\gamma\lambda\nu\xi}\mathcal{V}^{\lambda}\mathcal{V}^{\xi}
=\frac{1}{2}\eta_{\gamma\lambda\alpha\nu}\mathcal{R}^{\alpha\nu}_{\nu\xi}\mathcal{V}^{\lambda}\mathcal{V}^{\xi},
\\\label{31}
\mathcal{X}_{\gamma\nu}&=&^{*}\mathcal{R}^{*}_{\gamma\lambda\nu\xi}\mathcal{V}^{\lambda}\mathcal{V}^{\xi}
=\frac{1}{2}\eta^{\alpha\nu}_{\gamma\nu}\mathcal{R}^{*}_{\alpha\nu\lambda\xi}\mathcal{V}^{\lambda}\mathcal{V}^{\xi},
\end{eqnarray}
where $\mathcal{R}^{*}_{\gamma\nu\lambda\xi}=\frac{1}{2}\eta
_{\alpha\nu\lambda\xi}\mathcal{R}^{\alpha\nu}_{\gamma\nu}$. Using
above relations, we have
\begin{eqnarray}\label{32}
\mathcal{X}_{\gamma\nu}&=&\frac{1}{3}\mathcal{X}_{T}h_{\gamma\nu}+\mathcal{X}_{TF}\bigg
(\chi_{\gamma}\chi_{\nu}-\frac{1}{3} h_{\gamma\nu}\bigg),
\\\label{33}
\mathcal{Y}_{\gamma\nu}&=&\frac{1}{3}\mathcal{Y}_{T}h_{\gamma\nu}+\mathcal{Y}_{TF}\bigg
(\chi_{\gamma}\chi_{\nu}-\frac{1}{3} h_{\gamma\nu}\bigg),
\\\label{34}
\mathcal{Z}_{\gamma\nu}&=&\frac{1}{3}\mathcal{Z}_{T}h_{\gamma\nu}+\mathcal{Z}_{TF}\bigg
(\chi_{\gamma}\chi_{\nu}-\frac{1}{3} h_{\gamma\nu}\bigg).
\end{eqnarray}
The trace-free components of these functions are computed as
\begin{eqnarray}\label{35}
\mathcal{X}_{T}&=&8\pi\bigg(\rho+\frac{\mathcal{T}^{(\mathcal{GB})}_{00}}
{\mathcal{J}^2}\bigg),
\\\label{36}
\mathcal{X}_{TF}&=&-4\pi\bigg(\Pi+\frac{\mathcal{T}^{(\mathcal{GB})}_{11}}{\mathcal{K}^2}
-\frac{\mathcal{T}^{(\mathcal{GB})}_{22}}{\mathcal{L}^2}\bigg)-\xi,
\\\label{37}
\mathcal{Y}_{T}&=&4\pi\bigg(\rho+3\mathcal{P}_{r}-2\Pi+\frac{\mathcal{T}^{(\mathcal{GB})}_{00}}
{\mathcal{J}^2}+\frac{\mathcal{T}^{(\mathcal{GB})}_{11}}{\mathcal{K}^2}+\frac{2\mathcal{T}^{
(\mathcal{GB})}_{22}}{\mathcal{L}^2}\bigg),
\\\label{38}
\mathcal{Y}_{TF}&=&\xi-4\pi\bigg(\Pi+\frac{\mathcal{T}^{(\mathcal{GB})}_{11}}
{\mathcal{K}^2}-\frac{\mathcal{T}^{(\mathcal{GB})}_{22}}{\mathcal{L}^2}\bigg).
\end{eqnarray}
We compute only the electric component of the Weyl tensor and derived the scalar $Y_{TF}$ from it. Using Eqs.\eqref{17} and \eqref{28} with field equations, we have
\begin{equation}\label{39}
\frac{3m}{\mathcal{L}^{3}}=4\pi\bigg(\rho-\Pi+\frac{\mathcal{T}^{(\mathcal{GB})}_{00}}{\mathcal{J}
^2}-\frac{\mathcal{T}^{(\mathcal{GB})}_{11}}{\mathcal{K}^2}+\frac{\mathcal{T}^{(\mathcal{GB})}_{22}}
{\mathcal{L}^2}\bigg)-\xi-\frac{1}{8\mathcal{L}^2}.
\end{equation}
Using the preceding equation, the value of $Y_{TF}$ and $X_{TF}$
turn out to be
\begin{eqnarray}\nonumber
Y_{TF}&=&\frac{4\pi}{\mathcal{L}^{3}}\int^{r}_{0}\mathcal{L}^{3}\bigg[D_{\mathcal{L}}\rho
-\frac{3}{\mathcal{L}}\bigg\{\frac{\mathcal{T}^{
(\mathcal{GB})}_{00}}{\mathcal{J}^2}+\frac{\mathcal{U}}{E}\bigg(q
-\frac{\mathcal{T}^{(\mathcal{GB})}_{01}}{\mathcal{J}\mathcal{K}}\bigg)\bigg\}\bigg]
\mathcal{L}^{'}dr-\frac{1}{2\mathcal{L}^2}
\\\label{40}
&-&8\pi\bigg(\Pi+\frac{\mathcal{T}^{(\mathcal{GB})}_{11}}{\mathcal{K}^2}
-\frac{\mathcal{T}^{(\mathcal{GB})}_{22}}{\mathcal{L}^2}
-\frac{\mathcal{T}^{(\mathcal{GB})}_{00}}{2\mathcal{J}^2}\bigg),
\\\nonumber
X_{TF}&=&-\frac{4\pi}{\mathcal{L}^{3}}\int^{r}_{0}\mathcal{L}^{3}\bigg[D_{\mathcal{L}}\rho
-\frac{3}{\mathcal{L}}\bigg\{\frac{\mathcal{T}^{(\mathcal{GB}
)}_{00}}{\mathcal{J}^2}+\frac{\mathcal{U}}{E}\bigg(q-\frac{\mathcal{T}^{(\mathcal{GB})}_{01}}
{\mathcal{J}\mathcal{K}}\bigg)\bigg\}\bigg]\mathcal{L}^{'}dr
\\\label{41}
&-&4\pi\frac{\mathcal{T}^{(\mathcal{G})}_{00}}{\mathcal{J}^2}+\frac{1}{2\mathcal{L}^2}.
\end{eqnarray}
A differential equation may be developed for the scalar function and
inhomogeneous energy density as \cite{37}
\begin{equation}\label{42}
\bigg\{X_{TF}+4\pi\bigg(\rho+\frac{\mathcal{T}^{(\mathcal{GB})}_{00}}
{\mathcal{J}^2}\bigg)\bigg\}'=-X_{TF}\frac{3\mathcal{L}'}{\mathcal{L}}
+4\pi\bigg(q-\frac{\mathcal{T}^{(\mathcal{GB})}_{01}}{\mathcal{J}\mathcal{K}}\bigg)
\bigg(\Theta-\sigma\bigg)\mathcal{K}+\frac{\mathcal{L}'}{2\mathcal{L}^3}.
\end{equation}
In the non-dissipative context, the previously mentioned formula is
\begin{equation}\label{43}
X_{TF}=0\Leftrightarrow
4\pi\bigg(\rho+\frac{\mathcal{T}^{(\mathcal{GB})}_{00}}{\mathcal{J}^2}\bigg)'
=\frac{\mathcal{L}'}{2\mathcal{L}^3}
-4\pi\frac{\mathcal{T}^{(\mathcal{GB})}_{01}}{\mathcal{J}}\bigg(\Theta
-\sigma\bigg).
\end{equation}
In the dissipative scenario, Eq.(\ref{42}) is expressed as
\begin{equation}\label{44}
X_{TF}=0\Leftrightarrow\bigg(\rho+\frac{\mathcal{T}^{(\mathcal{GB})}_{00}}{\mathcal{J}^2}\bigg)'
=\bigg(q-\frac{\mathcal{T}^{(\mathcal{GB})}_{01}}{\mathcal{J}\mathcal{K}}\bigg)\bigg(\Theta-\sigma\bigg)\mathcal{K}
+\frac{\mathcal{L}'}{8\pi \mathcal{L}^3}.
\end{equation}
The result clarifies the significance of $X_{TF}$ for the analysis
of energy density inhomogeneity.

\subsection{Study of Junction Constraints}

The Darmois junction conditions are mathematical conditions which
ensure that a spacetime manifold is smoothly joined across a
hypersurface. These conditions are crucial when dealing with
situations like phase transitions in the early universe, matching
interior and exterior solutions of stars and black holes, or
describing thin shells and domain walls in spacetime. The first
condition ensures the continuity of the metric, preventing physical
singularities. This condition is used to describe how different
regions of spacetime can be smoothly connected at a boundary. This
constraint provides a way to match two different solutions of the
field equations across a hypersurface, which is often used to model
situations where one region demonstrates an interior solution and
the other region manifests an exterior solution. This is important
to ensure a smooth transition between the interior and exterior
solutions, maintaining the integrity of the spacetime geometry. The
second condition ensures the smoothness of the extrinsic curvature,
determining whether a surface layer of matter exists. If the second
condition is violated, a thin shell with a well-defined
stress-energy tensor appears, playing a key role in various
astrophysical and cosmological models.

We consider external spacetime as
\begin{equation}\label{45}
ds^{2}=\frac{2M(\upsilon)}{r}d\upsilon^{2}-2drd\upsilon+r^{2}(d\theta^{2}+\alpha^2dz^{2}),
\end{equation}
where $\upsilon$ defines the delayed time and $M(\upsilon)$
signifies the entire mass. The Darmois junction conditions yield
\begin{eqnarray}\label{46}
&&m(t,r)-M(\upsilon)\approx\frac{\mathcal{L}}{2}, \\\label{47}
&&q_{\Sigma} \approx
\mathcal{P}_{r}+\frac{\mathcal{T}^{(\mathcal{GB})}_{11}}{\mathcal{K}^2}-\frac{\mathcal{T}
^{(\mathcal{GB})}_{01}}{\mathcal{J}\mathcal{K}}.
\end{eqnarray}
The Eq.\eqref{46} expresses the relationship between the effective
gravitational mass $(m(t,r))$ of the interior fluid and the mass
function $(M(\upsilon))$ of the exterior Vaidya spacetime. This
ensures that the total mass of the system is correctly matched at
the boundary. This equation guarantees the smooth outflow of energy
across the boundary. The term  $q_{\Sigma}$ in Eq.\eqref{47}
demonstrates the radial heat flux which is evaluated at the boundary
$(\Sigma)$. It quantifies the heat flow across the boundary surface
between the interior and exterior spacetimes. The expression
$q_{\Sigma}\approx
\mathcal{P}_{r}+\frac{\mathcal{T}^{(\mathcal{GB})}_{11}}{\mathcal{K}^2}-\frac{\mathcal{T}
^{(\mathcal{GB})}_{01}}{\mathcal{J}\mathcal{K}}$ relates the heat
flux at the boundary to the radial pressure of the fluid, a
gravitational contribution from the modified energy-momentum tensor
$(\mathcal{T}^{(\mathcal{GB})}_{11})$ and
$(\mathcal{T}^{(\mathcal{GB})}_{01})$ defines an additional flux
term due to modified gravity effects. This equation plays a key role
in ensuring that the junction conditions are satisfied at the
boundary between the interior fluid configuration and the exterior
Vaidya radiating solution. Thus, these equations ensure that the
interior solution describing a self-gravitating fluid is
consistently connected to the exterior radiating solution,
preserving the physical consistency of the model.

\section{Examination of Complexity Factor}

The definition that measures the complexity of a dynamical system is
more generalized than that of a static system as it encounters two
extra elements. In the static scenario, the fluid characteristics
are important, however in the non-static scenario, the complexity of
the system's structure and the patterns of evolution play a
significant role. Nevertheless, in the latter scenario, the most
elementary patterns are also evaluated to assess the complexity of
evolutionary patterns. We choose $Y_{TF}$ as the complexity factor
since it encompasses all components that contribute to a system's
complexity. In the present situation, we identified $Y_{TF}$ as the
most appropriate scalar for analyzing the components that induce
problems inside a system. It further includes the impacts of dark
source words. Consequently, it also quantifies the geometric
variances in a system. We now investigate the complexity when
combined with $\mathcal{GB}$ manipulates for the dynamical systems.

We can find out two fundamental evolutionary techniques based on
basic principles, i.e., homogeneous expansion, characterized by a
constant expansion rate $({\Theta}'=0)$ and homologous evolution.

\subsection{Evolution Corresponding to Homologous and Homogeneous}

A homogeneous expansion defines a recognizable evolutionary pattern that is described as simple. For homologous evolution, Eq.(\ref{16}) is expressed as
\begin{equation}\label{48}
D_{\mathcal{L}}\bigg(\frac{\mathcal{U}}{\mathcal{L}}\bigg)=\frac{4\pi}{E}\bigg(q-\frac{\mathcal{T}^
{(\mathcal{GB})}_{01}}{\mathcal{J}\mathcal{K}}\bigg)+\frac{\sigma}{\mathcal{L}}.
\end{equation}
By integrating the previous equation, we obtain
\begin{equation}\label{49}
\mathcal{U}=a(t)\mathcal{L}+\mathcal{L}\int^{r}_{0}\bigg[\frac{4\pi}{E}\bigg(q-\frac{\mathcal{T}^
{(\mathcal{GB})}_{01}}{\mathcal{J}\mathcal{K}}\bigg)+\frac{\sigma}{\mathcal{L}}\bigg]\mathcal{L}'dr,
\end{equation}
Substituting the value of the integration function $a(t)$ into the
previous equation, we have
\begin{equation}\label{50}
\mathcal{U}=\frac{\mathcal{U}_{\Sigma}}{\mathcal{L}_{\Sigma}}\mathcal{L}-\mathcal{L}\int^{r_{\Sigma}}_{r}
\bigg[\frac{4\pi}{E}\bigg(q-\frac{\mathcal{T}^{(\mathcal{GB})}_{01}}
{\mathcal{J}\mathcal{K}}\bigg)+\frac{\sigma}{\mathcal{L}}\bigg]\mathcal{L}'dr.
\end{equation}
From the previous Eqs.(\ref{49}) and (\ref{50}), we derive that
$\mathcal{U}=\mathcal{L}$, which exhibits a characteristic of
homologous evolution. For two areal radii $\mathcal{L}_{I}$ and
$\mathcal{L}_{II}$ then
\begin{equation}\label{51}
\frac{\mathcal{L}_{I}}{\mathcal{L}_{II}}=constant.
\end{equation}
This equation determines that during the progression of fluid distribution, the evolutionary pattern corresponding to the homologous criteria is the most direct. Consequently, for the
identical analysis
\begin{equation}\label{52}
\mathcal{U}=a(t)\mathcal{L}, \quad a(t)\equiv
\frac{\mathcal{U}_{\Sigma}}{\mathcal{L}_{\Sigma}}.
\end{equation}
Consequently, $\mathcal{L}$ is a separable function; therefore,
\begin{equation}\label{53}
\mathcal{L}=\mathcal{L}_{1}(t)\mathcal{L}_{2}(r).
\end{equation}
By employing Eqs.\eqref{51} \eqref{53} in \eqref{19}, the homologous condition is established as
\begin{equation}\label{54}
\frac{4\pi\mathcal{K}}{\mathcal{L}'}\bigg(q-\frac{\mathcal{T}^{(\mathcal{GB})}_{01}}{\mathcal{J}\mathcal{K}}\bigg)
+\frac{\sigma}{\mathcal{L}}=0.
\end{equation}

A homogeneous expansion is an alternative evolutionary creation that
may be prohibited as a simplistic method, Eq.(\ref{19}) produces
\begin{equation}\label{55}
4\pi\bigg(q-\frac{\mathcal{T}^{(\mathcal{GB})}_{01}}{\mathcal{J}\mathcal{K}}\bigg)=-\frac{\mathcal{L}'}{\mathcal{K}}
\bigg[\frac{1}{3}D_{\mathcal{L}}(\sigma)+\frac{\sigma}{\mathcal{L}}\bigg].
\end{equation}
Using the homologous condition we acquire the conclusion
$D_{\mathcal{L}}(\sigma)=0$, which indicates that $\sigma=0$,
denoting the lack of dissipation. The fluid is homogeneous as
indicated by Eq.(\ref{50}).

\section{Study of Kinematical Parameters}

The homologous condition depicted in Eq.(\ref{54}) is expressed as
\begin{equation}\label{56}
4\pi\mathcal{K}\bigg(q-\frac{\mathcal{T}^{(\mathcal{GB})}_{01}}{\mathcal{J}\mathcal{K}}\bigg)=-\frac{\mathcal{L}'
\sigma}{\mathcal{L}}.
\end{equation}
Substituting the preceding value into Eq.(\ref{20}), we have
\begin{equation}\label{57}
(\Theta-\sigma)'=0.
\end{equation}
By substituting the values of $\sigma$ and $\Theta$ , we obtain
\begin{equation}\label{58}
(\Theta-\sigma)'=\bigg(\frac{3\dot{\mathcal{L}}}{\mathcal{J}\mathcal{L}}\bigg)'=0.
\end{equation}
Using Eq.(\ref{53}), we derive
\begin{equation}\label{59}
\mathcal{J}'=0.
\end{equation}
This geodesic conditions necessitate a homologous fluid. Indeed, we obtain $\mathcal{J}=1$ from this condition, which is given by
\begin{equation}\label{60}
\Theta-\sigma=\frac{3\dot{\mathcal{L}}}{\mathcal{L}}.
\end{equation}
We obtain $(\Theta-\sigma)'=0$ for $\mathcal{L}\sim{r}$ by assessing
the earlier equation in the center. We identify Eq.(\ref{60}) around
the centre by computing many derivatives with regard to $r$ as
\begin{equation}\label{61}
\frac{{\partial}^n(\Theta-\sigma)}{\partial{r^n}}=0, \quad n>0,
\end{equation}
where homologous factors are interdependent. This finding is compatible with \cite{38}. The shear-free condition necessitates the corresponding condition in this non-dissipative case. In the absence of dissipation, the homogeneous growth as shown by Eq.(\ref{55}) reveals
\begin{equation}\label{62}
\frac{\sigma^{'}}{\sigma}=-\frac{3\mathcal{L}'}{\mathcal{L}},
\end{equation}
integration of above equation, obtains the following form
\begin{equation}\label{63}
\sigma=\frac{b(t)}{\mathcal{L}^{3}},
\end{equation}
where $b(t)$ is an arbitrary integration function. At the center,
$\mathcal{L}$ equals to zero when $r=0$, which indicates that
$b(t)=0$. Hence $\sigma=0$. If $\sigma=0$ in Eq.(\ref{20}), then
${\Theta}'=0$ ultimately.
\begin{equation}\label{64}
\sigma=0\Leftrightarrow \mathcal{U}\sim
\mathcal{L}\Leftrightarrow\Theta^{'}=0.
\end{equation}
This finding is compatible with \cite{39}. It is important to highlight that the scalar function
$Y_{TF}=0$ if and only if a shear-free geodesic fluid maintains its geodesic and shear-free properties throughout its history \cite{40}. Consequently, if $Y_{TF}=0$ and the fluid follows a geodesic path, a system that begins its evolution from the outset will stay
shear-free. By inserting ${\Theta}'=0$, Eq.(\ref{8}) appears as
\begin{equation}\label{65}
\sigma^{'}+\frac{3\sigma\mathcal{L}'}{\mathcal{L}}=12\pi\frac{\mathcal{T}^{(\mathcal{GB})}_{01}}
{\mathcal{J}}.
\end{equation}
The integration of this equation gives
\begin{equation}\label{66}
\sigma=\frac{12\pi}{\mathcal{L}^3}\int\frac{\mathcal{T}^{(\mathcal{GB})}_{01}\mathcal{L}^3}
{\mathcal{J}}dr+\frac{b(t)}{\mathcal{L}^{3}}.
\end{equation}
The homologous condition contradicts the previous assertion unless we assume $\sigma=0$. It discusses the challenges of concurrently applying homogeneous and homologous expansion requirements.

\section{Dynamical Factors}

In the dynamically setting, the fluid adheres to a geodesic under analogous conditions. Therefore, under this condition, Eq.\eqref{eq:A9} is articulated as
\begin{equation}\label{67}
D_{\tau}\mathcal{U}=\frac{1}{8\mathcal{L}}-4\pi\mathcal{L}\bigg(\mathcal{P}_{r}+\frac{\mathcal{T}^{(
\mathcal{GB})}_{11}}{\mathcal{K}^2}\bigg)-\frac{m}{\mathcal{L}_{1}^{2}}.
\end{equation}
Applying the scalar function $Y_{TF}$ in the prior equation gives
\begin{equation}\label{68}
3\frac{D_{\tau}\mathcal{U}}{\mathcal{L}}=-4\pi\bigg(\rho+3\mathcal{P}_{r}-2\Pi+\frac{\mathcal{T}^{(\mathcal{GB}
)}_{00}}{\mathcal{J}^2}+\frac{\mathcal{T}^{(\mathcal{GB})}_{11}}{\mathcal{K}^2}+\frac{2\mathcal{T}
^{(\mathcal{GB})}_{22}}{\mathcal{L}^2}\bigg)+Y_{TF}+\frac{1}{2\mathcal{L}^2}.
\end{equation}
The mass function, combined with the $\mathcal{GB}$ field equations and
Eq.(\ref{28}), gives
\begin{equation}\label{69}
4\pi\bigg(\rho+3P_{r}-2\Pi+\frac{\mathcal{T}^{(\mathcal{G})}_{00}}{\mathcal{J}^2}
+\frac{\mathcal{T}^{(\mathcal{G})}_{11}}{\mathcal{K}^2}+\frac{2\mathcal{T}^
{(\mathcal{G})}_{22}}{\mathcal{L}^2}\bigg)=-\frac{2\ddot{\mathcal{L}}}{\mathcal{L}}
-\frac{\ddot{\mathcal{K}}}{\mathcal{K}}.
\end{equation}
using Eq.(\ref{18}), we obtain
\begin{equation}\label{70}
3\frac{D_{\tau}\mathcal{U}}{\mathcal{L}}=\frac{3\ddot{\mathcal{L}}}{\mathcal{L}}.
\end{equation}
Substituting the value from the prior equation into Eq.(\ref{68}), we attain
\begin{equation}\label{71}
\frac{\ddot{\mathcal{L}}}{\mathcal{L}}-\frac{\ddot{\mathcal{K}}}{\mathcal{K}}-\frac{1}
{2\mathcal{L}^2}=Y_{TF}.
\end{equation}
Assuming the fluid is homogeneous, and substituting Eq.(\ref{52})
into Eq.(\ref{68}), we get {\small{\begin{equation}\label{72}
3\bigg(\dot{a}(t)+a(t)\frac{\dot{\mathcal{L}}}{\mathcal{L}}\bigg)=-4\pi\bigg(\rho+3\mathcal{P}_{r}
-2\Pi+\frac{\mathcal{T}^{(\mathcal{GB})}_{00}}{\mathcal{J}^2}+\frac{\mathcal{T}^{(\mathcal{GB})}
_{11}}{\mathcal{K}^2}+\frac{2\mathcal{T}^{(\mathcal{GB})}_{22}}{\mathcal{L}^2}\bigg)+Y_{TF}
+\frac{1}{2\mathcal{L}^2}.
\end{equation}}}
The integration of Eq.(\ref{71}) for the non-complex system, define by a zero complexity factor as
\begin{equation}\label{73}
\mathcal{K}=\mathcal{L}_{1}(t)\bigg[k_{1}(r)\int\bigg(\frac{1}{\mathcal{L}^{2}_{1}(t)}e^{\frac{-1}
{\mathcal{L}^{2}_{2}(r)}\int\frac{dt}{\mathcal{L}_{1}(t)\dot{\mathcal{L}}_{1}(t)}}\bigg)dt+k_{2}
(r)\bigg],
\end{equation}
where $k_{1}(r)$ and $k_{2}(r)$ are integration functions. The above
equation may be simply expressed as
\begin{equation}\label{74}
\mathcal{K}=\mathcal{L}_{1}(t)\mathcal{L}_{2}^{'}(r)\bigg[\tilde{k}_{1}(r)\int\bigg(\frac{1}
{\mathcal{L}^{2}_{1}(t)}e^{\frac{-1}{\mathcal{L}^{2}_{2}(r)}\int\frac{dt}{\mathcal{L}_{1}(t)
\dot{\mathcal{L}}_{1}(t)}}\bigg)dt+\tilde{k}_{2}(r)\bigg].
\end{equation}
The values of $k_{1}(r)$ and $k_{2}(r)$ are $\mathcal{L}'_{2}(r)\tilde{k}_{1}(r)$ and
$\mathcal{L}'_{2}(r)\tilde{k}_{2}(r)$, by using these values, we introduce new variable
\begin{equation}\label{75}
\mathcal{S}=\tilde{k}_{1}(r)\int\bigg(\frac{1}{\mathcal{L}^{2}_{1}(t)}e^{\frac{-1}{\mathcal{L}^{2}_{2}(r)}
\int\frac{dt}{\mathcal{L}_{1}(t)\dot{\mathcal{L}}_{1}(t)}}
\bigg)dt+\tilde{k}_{2}(r),
\end{equation}
so
\begin{equation}\label{76}
\mathcal{K}=\mathcal{S}\mathcal{L}',
\end{equation}
where $\mathcal{L}'=\mathcal{L}_{1}(t)\mathcal{L}'_{2}(r)$.

\subsection{The Non-Dissipative Case}

For a non-dissipative fluid, by implementing the homologous condition and Eq.(\ref{15}) into Eq.(\ref{64}), we obtain
\begin{equation}\label{77}
\frac{\ddot{\mathcal{L}}}{\mathcal{L}}-\frac{\ddot{\mathcal{K}}}{\mathcal{K}}=0
\Rightarrow Y_{TF}=\frac{-1}{2\mathcal{L}^2}.
\end{equation}
The fluid is devoid of shear; hence, we derive the criteria from
Eqs.(\ref{15}) and (\ref{73}) as follows
\begin{equation}\label{78}
{k}_{1}(r)=0 \Rightarrow \mathcal{K}={\mathcal{L}}_{1}(t){k}_{2}(r)={\mathcal{L}}_{1}(t)
{\mathcal{L}'}_{2}(r)\tilde{k}_{2}(r).
\end{equation}
By applying the criteria of the prior equations and parameterized
$r$, we get $\mathcal{S}=1$, signifying this
$\mathcal{K}=\mathcal{L}'$ \cite{41}. The field equations rebuilt as
\begin{equation}\label{79}
4\pi\bigg(q-\frac{\mathcal{T}^{(\mathcal{GB})}_{01}}{\mathcal{S}\mathcal{L}'}\bigg)=
-\frac{\dot{\mathcal{S}}}{\mathcal{S}^{2}\mathcal{L}},
\end{equation}
\begin{equation}\label{80}
8\pi\bigg({\mathcal{P}_{r}}-\mathcal{P}_{\bot}+\frac{\mathcal{T}^{(\mathcal{GB})}_{11}}{S^2{\mathcal{L}'}^2}
-\frac{\mathcal{T}^{(\mathcal{G})}_{22}}{\mathcal{L}^2}\bigg)=\frac{\dot{\mathcal{S}}\dot{\mathcal{L}}}
{\mathcal{S}\mathcal{L}}+\frac{1}{\mathcal{S}^{2}\mathcal{L}^{2}}\bigg(\frac{\mathcal{S}'\mathcal{L}}{\mathcal{S}\mathcal{L}'}+1\bigg).
\end{equation}
This equation attains a value of zero when $\mathcal{S}=1$. This
system undoubtedly includes the most essential arrangement.
Homogeneous and homologous expansion are inherently interconnected
in a non-dissipative system.

\subsection{The Dissipative Condition}

For this scenario, we employ Eqs.(\ref{15}) and (\ref{71}) as follows
\begin{equation}\label{81}
\dot{\sigma}=-Y_{TF}-\frac{1}{2\mathcal{L}^2}+\bigg(\frac{\dot{\mathcal{L}}}{\mathcal{L}}\bigg)^{2}
-\bigg(\frac{\dot{\mathcal{K}}}{\mathcal{K}}\bigg)^{2}.
\end{equation}
Calculating the time derivative of Eq.(\ref{56}) modifies the previous equation as
\begin{equation}\label{82}
\frac{\mathcal{L}'}{\mathcal{L}}\bigg(Y_{TF}+\frac{1}{2\mathcal{L}^2}\bigg)=4\pi\mathcal{K}
\bigg(q-\frac{\mathcal{T}^{(\mathcal{GB})}_{01}}{\mathcal{K}}\bigg)\bigg(\frac{2\dot{\mathcal
{K}}}{\mathcal{K}}+\frac{\dot{\mathcal{L}}}{\mathcal{L}}+\frac{\frac{\partial}{\partial
t}
(q-\frac{\mathcal{T}^{(\mathcal{GB})}_{01}}{\mathcal{K}})}{(q-\frac{\mathcal{T}^{(\mathcal{GB}
)}_{01}}{\mathcal{K}})}\bigg),
\end{equation}
By implementing the vanishing complexity constraint, we have
\begin{equation}\label{83}
q-\frac{\mathcal{T}^{(\mathcal{GB})}_{01}}{\mathcal{K}}=\frac{1}{8\pi
\mathcal{K}^2\mathcal{L}}
\int\frac{\mathcal{L}'\mathcal{K}}{\mathcal{L}^2}dt+\frac{f(r)}{\mathcal{K}^{2}\mathcal{L}}.
\end{equation}
Also,
\begin{equation}\label{84}
\frac{\partial}{\partial
t}\bigg(q-\frac{\mathcal{T}^{(\mathcal{GB})}_{01}}{\mathcal{K}}\bigg)
=-\bigg(q-\frac{\mathcal{T}^{(\mathcal{GB})}_{01}}{\mathcal{K}}\bigg)\bigg(\Theta+\sigma\bigg)
+\frac{\mathcal{L}'}{8\pi \mathcal{K}\mathcal{L}^3}.
\end{equation}
The techniques described in \cite{42} can be utilized to get details of this nature. The lack of transient events (stationary state) is presently regarded as the most critical dissipative regime in a dissipative process \cite{43}. Considering the situation is stable, we deduce
\begin{equation}\label{85}
q-\frac{\mathcal{T}^{(\mathcal{GB})}_{01}}{\mathcal{K}}=-\frac{\kappa\mathcal{T}'}{\mathcal{K}}.
\end{equation}
Substituting the value from Eq.(\ref{83}) into the preceding equation, we have
\begin{equation}\label{86}
\mathcal{T}'=-\frac{1}{8\pi \kappa
\mathcal{K}\mathcal{L}}\int\frac{\mathcal{L}'\mathcal{K}}{\mathcal{L}^2}dt-\frac
{f(r)}{\kappa \mathcal{K}\mathcal{L}}.
\end{equation}
At this stage, we can neither uphold the assumption regarding the disappearance of the relaxation time as a sign of minimal complexity inside the dissipative regime, nor can we demonstrate the existence of such exact solutions.

\section{The Stability of Non-Complex Structures}

We analyze the advancement of the scalar function $X_{TF}$ given in
Eq.\eqref{eq:A10} as
\begin{align}\nonumber
&\frac{3\dot{\mathcal{L}}}{\mathcal{L}}Y_{TF}+8\pi\frac{\partial}{\partial
t}\bigg(\Pi+\frac{\mathcal{T}^{(\mathcal{GB})}_{11}}{\mathcal{K}^2}-\frac{\mathcal{T}
^{(\mathcal{GB})}_{22}}{\mathcal{L}^2}\bigg)+4\pi\sigma\bigg(\rho+\mathcal{P}_{r}
+\frac{\mathcal{T}^{(\mathcal{GB})}_{00}}{\mathcal{J}^2}+\frac{\mathcal{T}^
{(\mathcal{GB})}_{11}}{\mathcal{K}^2}\bigg)+\\\nonumber&16\pi\bigg(\Pi
+\frac{\mathcal{T}^{(\mathcal{GB})}_{11}}{\mathcal{K}^2}-\frac{\mathcal{T}^{
(\mathcal{GB})}_{22}}{\mathcal{L}^2}\bigg)\frac{\dot{\mathcal{L}}}{\mathcal{L}}+\frac{4\pi}
{\mathcal{K}}\bigg[\frac{\partial}{\partial
r}\bigg(q-\frac{\mathcal{T}
^{(\mathcal{GB})}_{01}}{\mathcal{K}}\bigg)-\bigg(q-\frac{\mathcal{T}^{(\mathcal{GB})}_{01}}
{\mathcal{K}}\bigg)\frac{\mathcal{L}'}{\mathcal{L}}\bigg)\bigg]
\\\label{87}&+\dot{Y}_{TF}+\frac{\dot{\mathcal{L}}}{2\mathcal{L}^3}=0.
\end{align}
The aforementioned equation can be classified as
\begin{align}\nonumber
&\dot{Y}_{TF}+8\pi\frac{\partial}{\partial
t}\bigg(\Pi+\frac{\mathcal{T}^{(\mathcal{GB})}
_{11}}{\mathcal{K}^2}-\frac{\mathcal{T}^{(\mathcal{GB})}_{22}}{\mathcal{L}^2}\bigg)+\frac{4\pi}
{\mathcal{K}}\bigg[\frac{\mathcal{L}'\mathcal{T}^{(\mathcal{GB})}_{01}}{\mathcal{K}\mathcal{L}}-\frac{\partial}{\partial
r}\bigg(\frac{\mathcal{T}^{(\mathcal{GB})}_{01}}{\mathcal{K}}\bigg)\bigg]\\\label{88}
+&\frac{\dot{\mathcal{L}}}{2\mathcal{L}^3}+16\pi\bigg(\frac{\mathcal{T}^{(\mathcal{GB})}_{11}}
{\mathcal{K}^2}-\frac{\mathcal{T}^{(\mathcal{GB})}_{22}}{\mathcal{L}^2}\bigg)\frac{\dot{\mathcal
{L}}}{\mathcal{L}}=0.
\end{align}
Taking derivative of  Eq.(\ref{40}) with respect to time and evaluating it at $t=0$, as well as differentiating Eq.(\ref{87}) and substituting $Y_{TF}=0$, the only viable solution is
\begin{align}\nonumber
&8\pi\frac{\partial^2}{{\partial{t}^2}}\bigg[\Pi+\frac{\mathcal{T}^{(\mathcal{GB})}_{11}}{\mathcal
{K}^2}-\frac{\mathcal{T}^{(\mathcal{GB})}_{22}}{\mathcal{L}^2}\bigg]+16\pi\bigg[\frac{\dot{\Pi}
\dot{\mathcal{L}}}{\mathcal{L}}-\frac{\partial}{\partial
t}\bigg\{\bigg(\frac{\mathcal{T}
^{(\mathcal{GB})}_{11}}{\mathcal{K}^2}-\frac{\mathcal{T}^{(\mathcal{GB})}_{22}}{\mathcal{L}^2}\bigg)
\frac{\dot{\mathcal{L}}}{\mathcal{L}}\bigg\}\bigg]\\\label{89}&+\ddot{Y}_{TF}+\frac{3\dot
{\mathcal{L}}\dot{Y}_{TF}}{\mathcal{L}}+4\pi\frac{\partial}{\partial
t}\bigg\{\frac{\mathcal{L}
'\mathcal{T}^{(\mathcal{GB})}_{01}}{\mathcal{K}^2\mathcal{L}}-\frac{1}{\mathcal{K}}\frac{\partial}
{\partial
r}\bigg(\frac{\mathcal{T}^{(\mathcal{GB})}_{01}}{\mathcal{K}}\bigg)\bigg\}
+\frac{\partial}{\partial
t}\bigg(\frac{\dot{\mathcal{L}}}{2\mathcal{L}^3}\bigg)=0.
\end{align}
Calculating the second derivative of Eq.(\ref{42}) gives
{\small\begin{equation}\label{90} 2\frac{\partial}{\partial
t}\bigg(\Pi+\frac{\mathcal{T}^{(\mathcal{GB})}_{11}}{\mathcal{K}^2}
-\frac{\mathcal{T}^{(\mathcal{GB})}_{22}}{\mathcal{L}^2}\bigg){\dot{\mathcal{L}}}-\frac{1}{4\pi
\mathcal{L}}\bigg[\frac{\ddot{\mathcal{L}}}{\mathcal{L}}-3\bigg(\frac{\dot{\mathcal{L}}}
{\mathcal{L}}\bigg)^2\bigg]=\frac{1}{\mathcal{L}^{2}}\int^{r}_{0}\bigg[\frac{\partial^2}
{{\partial{t}^2}}\bigg({\rho+\mathcal{T}^{(\mathcal{GB})}_{00}}\bigg)\bigg]'dr.
\end{equation}}
In general case, at the first stage of the dissipative system, we possess
\begin{align}\nonumber
&\dot{Y}_{TF}+8\pi\frac{\partial}{\partial
t}\bigg(\Pi+\frac{\mathcal{T}^{(\mathcal{GB})}
_{11}}{\mathcal{K}^2}-\frac{\mathcal{T}^{(\mathcal{GB})}_{22}}{\mathcal{L}^2}\bigg)+4\pi\sigma
\bigg(\rho+\mathcal{P}_{r}+\frac{\mathcal{T}^{(\mathcal{GB})}_{00}}{\mathcal{J}^2}+\frac{\mathcal{T}^
{(\mathcal{GB})}_{11}}{\mathcal{K}^2}\bigg)+\\\nonumber&16\pi\bigg(\Pi+\frac{\mathcal{T}^
{(\mathcal{GB})}_{11}}{\mathcal{K}^2}-\frac{\mathcal{T}^{(\mathcal{G})}_{22}}{\mathcal{L}^2}
\bigg)\frac{\dot{\mathcal{L}}}{\mathcal{L}}+\frac{4\pi}{\mathcal{K}}\bigg[\frac{\partial}
{\partial
r}\bigg(q-\frac{\mathcal{T}^{(\mathcal{G})}_{01}}{\mathcal{K}}\bigg)-\bigg(q-\frac
{\mathcal{T}^{(\mathcal{G})}_{01}}{\mathcal{K}}\bigg)\frac{\mathcal{L}'}{\mathcal{L}}\bigg)
\bigg]\\\label{91}&+\frac{\dot{\mathcal{L}}}{2\mathcal{L}^3}=0.
\end{align}
This discussion focusses solely on the influence of the efficient
terms of $\mathcal{GB}$ gravity.

\section{Discussion and Final Outcomes}

The examination of astrophysical objects is a fascinating phenomena
that inspires scientists to investigate the physical characteristics
of these objects. The physical characteristics like anisotropy,
energy density, stability/instability, and luminosity of stellar
structures have been extensively researched in theoretical work.
However, the topic of complexity has not been extensively addressed
for compact objects. The objective of this study is to examine the
impact of correction terms on the complexity factor of
time-dependent cylindrical symmetric spacetime in $f(\mathcal{G})$
theory. We have developed the modified field equations with an
anisotropic matter source using the hydrostatic equilibrium
condition. We have calculated the mass distribution function using
the C-energy framework. Moreover, we have explored the correlation
between the mass distribution functions and the physical parameters
with the Weyl tensor. We have analyzed the structural scalars and
determined the complexity factor corresponding to these scalars.

The complex nature of dynamical cylindrical configuration in the
framework of modified Gauss-Bonnet gravity has been examined. We
have examined two distinct interrelated attributes of the complexity
with the anisotropic matter configuration. We have examined both the
complexity of the fluid's structure and the complexity of the fluid
distribution's evolutionary pattern. We have selected the scalar
function $Y_{TF}$ as a measure of the fluid's structural complexity.
The complexity factor remains consistent with that of the non-static
case. In the non-dissipative instance, the homologous condition
involves the elimination of $Y_{TF}$. We have observed that the term
$Y_{TF}$ specifically encompasses the cylindrical structural effects
caused by density inhomogeneity, modified terms and anisotropic
stresses. Furthermore, the additional curvature elements of
$f(\mathcal{G})$ impose limitations on the gravitational structure,
preventing it from losing its homogenous state. We have focused on
the complexity of evolutionary patterns. Two alternatives emerge as
the most apparent candidates: the homologous condition and the
homogenous expansion. This indicates that the fluid is geodesic,
even in the most generic dissipative case. The geodesic flow clearly
shows one of the most fundamental patterns of advancement. The
contribution of modified terms, anisotropic stresses, and
inhomogeneous densities is quantified by the dynamical variable
$Y_{TF}$ in a specific sequence.

In the non-dissipative context, the circumstance indicates that
$Y_{TF}$ signifies that the most basic pattern efficiently predicts
the fundamental configuration of fluid distribution. In the
non-dissipative scenario, it results in a singular solution. Next,
we have addressed the issue of the stability of the vanishing
complexity factor condition. In the non-dissipative situation, it is
evident that this condition would spread over time which provides
that the pressure remains isotropic. In the dissipative scenario,
the circumstances are far more intricate and dissipative terms may
also cause the system to diverge from the requirement of vanishing
complexity factor. Our analysis of existing research reveals that
the complexity factor for cylindrical systems includes additional
elements due to the geometric distinctions of self-gravitating
systems, preventing it from reaching a value of zero in the simplest
evolutionary modes. The difficulty of non-static self-gravitating
systems has been well analyzed; however, further research is
required for dynamical systems. The inclusion of extra curvature
components leads to an increase in the complexity of the cylindrical
structure. The resulting equations have been derived to describe the
energy density, radial pressure, and tangential pressure behavior of
these particles, respectively. This set of differential equations
offer the solution given certain appropriate initial conditions to
enhance comprehension of the system.\\\\\\
\textbf{Acknowledgements}: M. Zeeshan Gul sincerely thanks Prof. Ni
Fei of the College of Transportation, Tongji University, China, for
her invaluable guidance and support. Her insightful mentorship
greatly shaped this research and this work reflects her lasting
impact on my academic journey.

\section*{Appendix A}

The additional curvature terms resulting from GB gravity are shown
below.
\begin{align}\tag{A1}\label{eq:A1}
\mathcal{T}^{(\mathcal{GB})}_{00}&=\frac{\mathcal{J}^2}{2}\bigg(\mathcal{G}f_{\mathcal{G}}
-f\bigg)-\mathcal{S}_{1}f^{''}_{\mathcal{G}}-\mathcal{S}_{2}f^{'}_{\mathcal{G}}-\mathcal{S}_{3}\dot{f_{\mathcal{G}}},
\\\tag{A2}\label{eq:A2}
\mathcal{T}^{(\mathcal{GB})}_{01}&=-\mathcal{S}_{4}\dot{f_{\mathcal{G}}}-\mathcal{S}_{5}f^{'}_{\mathcal{G}}-\mathcal{S}_{6}
\dot{f'_{\mathcal{G}}},
\\\tag{A3}\label{eq:A3}
\mathcal{T}^{(\mathcal{GB})}_{11}&=-\frac{\mathcal{K}^2}{2}\bigg(\mathcal{G}
f_{\mathcal{G}}
-f\bigg)-\mathcal{S}_{7}\dot{f_{\mathcal{G}}}-\mathcal{S}_{8}f^{'}_{\mathcal{G}}-\mathcal{S}_{9}\ddot{f_{\mathcal{G}}},
\\\tag{A4}\label{eq:A4} \mathcal{T}^{(\mathcal{GB})}_{22}&=-\frac{\mathcal{L}^2}{2}\bigg(\mathcal{G} f_{\mathcal{G}}-f\bigg)-\mathcal{S}_{10}\ddot{f_{\mathcal{G}}}-\mathcal{S}_{11}f^{''}_{\mathcal{G}}-\mathcal{S}_{12}\dot
{f_{\mathcal{G}}}-\mathcal{S}_{13}f^{'}_{\mathcal{G}}-\mathcal{S}_{14}\dot{f'_{\mathcal{G}}},
\end{align}
where
\begin{align}\nonumber
\mathcal{S}_{1}&=\frac{4\Big({\mathcal{L}'}^{2}{\mathcal{J}}^{2}-{\mathcal{J}}^{2}{\mathcal{K}}^{2}-
{\dot{\mathcal{L}}}^{2}{\mathcal{K}}^{2}\Big)}{{\mathcal{K}}^{4}{\mathcal{L}}^{2}},
\\\nonumber
\mathcal{S}_{2}&=\frac{4\Big({\mathcal{K}'}{\mathcal
{K}}^{2}{\mathcal{J}}{\dot{\mathcal{L}}}^{2}+2{\mathcal{L}'}{\mathcal{L}''}{\mathcal{K}}
{\mathcal{J}}^{3}-3{\mathcal{L}'}^{2}{\mathcal{K}'}{\mathcal{J}}^{3}+{\mathcal{K}'}{\mathcal{K}}
^{2}{\mathcal{J}}^{3}-2{\mathcal{L}'}{\dot{\mathcal{L}}}{\mathcal{K}}^{2}{\mathcal
{J}}{\dot{\mathcal{K}}}\Big)}{{\mathcal{J}}{\mathcal{K}}^{5}{\mathcal{L}}^{2}},
\\\nonumber
\mathcal{S}_{3}&=\frac{4\Big(3{\dot{\mathcal{K}}}{\mathcal{K}}^{2}{\dot{\mathcal{L}}}^{2}-{\mathcal{L}'}
^{2}{\dot{\mathcal{K}}}{\mathcal{J}}^{2}+{\mathcal{K}}^{2}{\dot{\mathcal{K}}}{\mathcal{J}}^{2}
-2{\mathcal{L}''}{\mathcal{J}}^{2}{\mathcal{K}}{\dot{\mathcal{L}}}+2{\dot{\mathcal{L}}}
{\mathcal{L}'}{\mathcal{K}'}{\mathcal{J}}^{2}\Big)}{{\mathcal{J}}^{2}{\mathcal{K}}^{3}
{\mathcal{L}}^{2}},
\\\nonumber
\mathcal{S}_{4}&=\frac{4\Big(2{\dot{\mathcal{L}}}{\mathcal{L}'}{\mathcal{J}}{\mathcal{K}}{\dot{\mathcal{K}}}
-{\mathcal{J}'}{\mathcal{J}}^{2}{\mathcal{L}'}^{2}-2{\mathcal{K}}^{2}{\dot{\mathcal{L}}}{\mathcal
{J}}{\dot{\mathcal{L}'}}+3{\mathcal{J}'}{\mathcal{K}}^{2}{\dot{\mathcal{L}}}^{2}+{\mathcal{J}'}
{\mathcal{K}}^{2}{\mathcal{J}}^{2}\Big)}{{\mathcal{K}}^{2}{\mathcal{J}}^{3}{\mathcal{L}}^{2}},
\\\nonumber
\mathcal{S}_{5}&=\frac{4\Big({\dot{\mathcal{L}}}^{2}{\mathcal{K}}^{2}{\dot{\mathcal{K}}}-2{\mathcal{J}'}
{\mathcal{J}}{\mathcal{K}}{\mathcal{L}'}{\dot{\mathcal{L}}}-3{\mathcal{J}}^{2}{\dot{\mathcal{K}}}
{\mathcal{L}'}^{2}+2{\mathcal{L}'}{\mathcal{J}}^{2}{\dot{\mathcal{L}}}^{'}{\mathcal{K}}+{\dot{
\mathcal{K}}}{\mathcal{K}}^{2}{\mathcal{J}}^{2}\Big)}{{\mathcal{K}}^{3}{\mathcal{J}}^{2}{\mathcal
{L}}^{2}},
\\\nonumber
\mathcal{S}_{6}&=\frac{4\Big({\mathcal{J}}^{2}{\mathcal{L}'}^{2}-{\mathcal{K}}^{2}{\dot{\mathcal{L}}}^{2}
-{\mathcal{K}}^{2}{\mathcal{J}}^{2}\Big)}{{\mathcal{K}}^{2}{\mathcal{J}}^{2}{\mathcal{L}}^{2}},
\\\nonumber
\mathcal{S}_{7}&=\frac{4\Big(2{\dot{\mathcal{L}}}{\mathcal{L}'}{\mathcal{J}'}{\mathcal{J}}^{2}{\mathcal
{K}}^{2}-{\mathcal{J}}^{2}{\dot{\mathcal{J}}}{\mathcal{K}}^{2}{\mathcal{L}'}^{2}-2{\mathcal{J}}
{\mathcal{K}}^{4}{\dot{\mathcal{L}}}{\ddot{\mathcal{L}}}+3{\dot{\mathcal{J}}}{\mathcal{K}}^{4}
{\dot{\mathcal{L}}}^{2}+{\dot{\mathcal{J}}}{\mathcal{J}}^{2}{\mathcal{K}}^{2}{\mathcal{L}}^{2}
\Big)}{{\mathcal{K}}^{2}{\mathcal{J}}^{5}{\mathcal{L}}^{2}},
\\\nonumber
\mathcal{S}_{8}&=\frac{4\Big({\dot{\mathcal{L}}}^{2}{\mathcal{J}'}{\mathcal{K}}^{2}-2{\dot{\mathcal{J}}}
{\mathcal{K}}^{2}{\mathcal{L}'}{\dot{\mathcal{L}}}-3{\mathcal{J}'}{\mathcal{J}}^{2}{\mathcal{L}
'}^{2}+2{\mathcal{J}}{\mathcal{K}}^{2}{\mathcal{L}'}{\ddot{\mathcal{L}}}+{\mathcal{J}'}{\mathcal
{J}}^{2}{\mathcal{K}}^{2}\Big)}{{\mathcal{K}}^{2}{\mathcal{J}}^{3}{\mathcal{L}}^{2}},
\\\nonumber
\mathcal{S}_{9}&=\frac{4\Big({\mathcal{L}'}^{2}{\mathcal{J}}^{2}{\mathcal{K}}^{2}-{\dot{\mathcal{L}}}^{2}
{\mathcal{K}}^{4}-{\mathcal{J}}^{2}{\mathcal{K}}^{4}\bigg)}{{\mathcal{K}}^{2}{\mathcal{J}}^{4}
{\mathcal{L}}^{2}},
\\\nonumber
\mathcal{S}_{10}&=\frac{4\Big({\mathcal{L}''}{\mathcal{J}}^{2}{\mathcal{K}}{\mathcal{L}}-{\dot{\mathcal
{K}}}{\dot{\mathcal{L}}}{\mathcal{K}}^{2}{\mathcal{L}}-{\mathcal{L}}{\mathcal{L}'}{\mathcal{K}'}
{\mathcal{J}}^{2}\Big)}{{\mathcal{J}}^{4}{\mathcal{K}}^{3}},
\\\nonumber
\mathcal{S}_{11}&=\frac{4\Big({\mathcal{L}}{\ddot{\mathcal{L}}}{\mathcal{J}}{\mathcal{K}}^{2}-{\mathcal
{J}'}{\mathcal{J}}^{2}{\mathcal{L}'}{\mathcal{L}}-{\mathcal{L}}{\dot{\mathcal{L}}}{\dot{\mathcal
{J}}}{\mathcal{K}}^{2}\Big)}{{\mathcal{J}}^{3}{\mathcal{K}}^{4}},
\\\nonumber
\mathcal{S}_{12}&=\frac{4}{{\mathcal{K}}^{3}{\mathcal{J}}^{5}}\Big({\dot{\mathcal{J}}}{\mathcal{J}}^{2}
{\mathcal{L}'}{\mathcal{L}}{\mathcal{K}'}-{\dot{\mathcal{K}}}{\mathcal{J}}^{2}{\mathcal{L}'}
{\mathcal{L}}{\mathcal{J}'}-{\mathcal{J}}{\mathcal{K}}^{2}{\dot{\mathcal{K}}}{\mathcal{L}}
{\ddot{\mathcal{L}}}-{\dot{\mathcal{J}}}{\mathcal{J}}^{2}{\mathcal{K}}{\mathcal{L}''}{\mathcal{L}}
\\\nonumber&
+2{\mathcal{J}'}{\mathcal{J}}^{2}{\mathcal{K}}{\mathcal{L}}{\dot{\mathcal{L}'}}-{\mathcal{J}}
{\mathcal{K}}^{2}{\ddot{\mathcal{K}}}{\mathcal{L}}{\dot{\mathcal{L}}}+{\mathcal{J}''}{\mathcal
{J}}^{2}{\mathcal{K}}{\mathcal{L}}{\dot{\mathcal{L}}}-2{\mathcal{J}}{\mathcal{J}'}^{2}{\mathcal
{K}}{\mathcal{L}}{\dot{\mathcal{L}}}
+3{\dot{\mathcal{J}}}{\dot{\mathcal{K}}}{\mathcal{K}}^{2}{\mathcal{L}}{\dot{\mathcal{L}}}
\\\nonumber&
-{\mathcal{J}'}{\mathcal{J}}^{2}{\mathcal{K}'}{\mathcal{L}}{\dot{\mathcal{L}}}\bigg),
\\\nonumber \mathcal{S}_{13}&=\frac{4}{{\mathcal{J}}^{3}{\mathcal{K}}^{5}}\bigg({\dot{\mathcal{J}}}{\mathcal{K}}^{2}
{\dot{\mathcal{L}}}{\mathcal{L}}{\mathcal{K}'}-{\dot{\mathcal{K}}}{\mathcal{K}}^{2}{\dot{
\mathcal{L}}}{\mathcal{L}}{\mathcal{J}'}-{\mathcal{J}}{\mathcal{K}}^{2}{\mathcal{K}'}{\mathcal
{L}}{\ddot{\mathcal{L}}}-{\mathcal{J}'}{\mathcal{J}}^{2}{\mathcal{K}}{\mathcal{L}''}{\mathcal{L}}
\\\nonumber&
+2{\mathcal{J}}{\mathcal{K}}^{2}{\dot{\mathcal{K}}}{\mathcal{L}}{\dot{\mathcal{L}'}}+{\mathcal
{J}}{\mathcal{K}}^{2}{\ddot{\mathcal{K}}}{\mathcal{K}}{\mathcal{L}'}-{\mathcal{J}''}{\mathcal
{J}}^{2}{\mathcal{K}}{\mathcal{L}}{\mathcal{L}'}-2{\mathcal{J}}{\dot{\mathcal{K}}}^{2}{\mathcal
{K}}{\mathcal{L}}{\mathcal{L}'}+3{\mathcal{J}'}{\mathcal{K}'}{C}^{2}{\mathcal{L}}{\mathcal{L}'}
\\\nonumber&
-{\dot{\mathcal{J}}}{\dot{\mathcal{K}}}{\mathcal{K}}^{2}{\mathcal{L}'}{\mathcal{L}}\bigg),
\\\nonumber
\mathcal{S}_{14}&=\frac{2\Big({\mathcal{L}}{\dot{\mathcal{K}}}{\mathcal{J}}{\mathcal{L}'}+{\mathcal{L}}
{\dot{\mathcal{L}}}{\mathcal{J}'}{\mathcal{K}}-{\mathcal{L}}{\dot{\mathcal{L}'}}{\mathcal{J}}
{\mathcal{K}}\Big)}{{\mathcal{J}}^{3}{\mathcal{K}}^{2}}.
\end{align}

\section*{Appendix B}

By using Eq.(\ref{7}), we obtain substantial formulas for the Bianchi identities as follows
\begin{align}\nonumber
\mathcal{T}^{\gamma\nu}_{;\nu}\mathcal{V}_{\gamma}&=-\frac{1}{\mathcal{J}}\bigg[{\dot{\rho}}+\bigg({\rho}
+\mathcal{P}_{r}\bigg)\frac{\dot{\mathcal{K}}}{\mathcal{K}}+2\bigg({\rho}+\mathcal{P}_{\bot}\bigg)\frac{\dot{\mathcal
{L}}}{\mathcal{L}}\bigg]
\\\tag{A5}\label{eq:A5}&
-\frac{1}{\mathcal{K}}\bigg[{q}^{'}+2{q}\bigg(\frac{\mathcal{J}'}{\mathcal{J}}+\frac{\mathcal{L}
'}{\mathcal{L}}\bigg)\bigg]+\mathcal{Z}{_1}=0,
\end{align}
here
\begin{align*}
\mathcal{Z}{_1}&=-\frac{1}{\mathcal{J}}\bigg[\frac{\partial}{\partial
t}\bigg(\frac{\mathcal{T}
_{00}^{\mathcal{(GB)}}}{\mathcal{J}^2}\bigg)+\frac{\mathcal{T}_{00}^{\mathcal{(GB)}}}{\mathcal{J}^2}
\bigg(\frac{2\dot{\mathcal{L}}}{\mathcal{L}}+\frac{\dot{\mathcal{K}}}{\mathcal{K}}\bigg)+\frac
{\dot{\mathcal{K}}\mathcal{T}_{11}^{\mathcal{(GB)}}}{\mathcal{K}^3}+\frac{2\dot{\mathcal{N}}\mathcal{T}
_{22}^{\mathcal{(GB)}}}{\mathcal{N}^3}\bigg]
\\\nonumber&
+\frac{1}{\mathcal{K}}\bigg[\bigg(\frac{\mathcal{T}_{01}^{\mathcal{(GB)}}}{\mathcal{J}\mathcal{K}}\bigg)'
+2\frac{\mathcal{T}_{01}^{\mathcal{(GB)}}}{\mathcal{J}\mathcal{K}}\bigg(\frac{\dot{\mathcal{L}}}{\mathcal{L}}
+\frac{\mathcal{J}'}{\mathcal{J}}\bigg)\bigg]
\end{align*}
\begin{align}\nonumber
\mathcal{T}^{\gamma\nu}_{;\nu}\chi_{\gamma}&=\frac{1}{\mathcal{K}}\bigg[{\mathcal{P}_{r}}^{'}+\bigg({\rho}
+\mathcal{P}_{r}\bigg)\frac{\mathcal{J}'}{\mathcal{J}}+2\bigg(\mathcal{P}_{r}-\mathcal{P}_{\bot}\bigg)\frac{\mathcal{L}'}
{\mathcal{L}}\bigg]
\\\tag{A6}\label{eq:A6}
&+\frac{1}{\mathcal{J}}\bigg[{\dot{q}}+2{q}\bigg(\frac{\dot{\mathcal{L}}}{\mathcal{L}}
+\frac{\dot{\mathcal{K}}}{\mathcal{K}}\bigg)\bigg]+\mathcal{Z}{_2}=0.
\end{align}
here
\begin{align*}
\mathcal{Z}{_2}&=\frac{1}{\mathcal{K}}\bigg[\bigg(\frac{\mathcal{T}_{11}^{\mathcal{(GB)}}}
{\mathcal{K}^2}\bigg)'+\frac{\mathcal{J}'\mathcal{T}_{00}^{\mathcal{(GB)}}}{\mathcal{J}^3}
-\frac{2\mathcal{L}'\mathcal{T}_{22}^{\mathcal{(GB)}}}{\mathcal{L}^3}+\frac{\mathcal{T}_{11}
^{\mathcal{(GB)}}}{\mathcal{K}^2}\bigg(\frac{2\mathcal{L}'}{\mathcal{L}}+\frac{\mathcal{J}'}
{\mathcal{J}}\bigg)\bigg]
\\\nonumber &
-\frac{1}{\mathcal{J}}\bigg[\frac{\partial}{\partial
t}\bigg(\frac{\mathcal{T}_{01}^{\mathcal{(GB)
}}}{\mathcal{J}\mathcal{K}}\bigg)+2\frac{\mathcal{T}_{01}^{\mathcal{(GB)}}}{\mathcal{J}\mathcal{K}}\bigg(\frac{\dot{\mathcal{L}}}{\mathcal
{L}}+\frac{\dot{\mathcal{K}}}{\mathcal{K}}\bigg)\bigg]
\end{align*}
Equations (\ref{9}), (\ref{17}), and (\ref{19}) are reformulated as follows
\begin{align}\nonumber
&D_{\tau}{\rho}+\frac{1}{3}\bigg(3{\rho}+\mathcal{P}_{r}+2\mathcal{P}_{\bot}\bigg)\Theta+\frac{2}{3}\bigg
(\mathcal{P}_{r}-\mathcal{P}_{\bot}\bigg)\sigma+2{q}\bigg(a+\frac{E}{\mathcal{L}}\bigg)
\\\tag{A7}\label{eq:A7}
&+ED_{\mathcal{L}}{q}-\mathcal{Z}=0,
\\\
&D_{\mathcal{T}}{q}+\frac{2}{3}{q}\bigg(2\Theta+\sigma\bigg)+\bigg({\rho}
+\mathcal{P}_{r}\bigg)a\bigg(\mathcal{P}_{r}-\mathcal{P}_{\bot}\bigg)\frac{2E}{\mathcal{L}}
\\\tag{A8}\label{eq:A8}
&+ED_{\mathcal{L}}{\mathcal{P}_{r}}+\mathcal{Z}=0.
\end{align}
The previously mentioned equation may be further simplified by
employing the mass function, as indicated in Eq.(\ref{9}) and
(\ref{19}), resulting in
\begin{align}\tag{A9}\label{eq:A9}
D_{\tau}\mathcal{U}=-\frac{m}{\mathcal{L}^2}-4\pi{\mathcal{P}_{r}}\mathcal{L}+Ea-\frac{4\pi
\mathcal{L}\mathcal{T}_{11}^{(\mathcal{GB})}}{\mathcal{K}^2}.
\end{align}
Applying \eqref{eq:A7}, we can derive the scalar function $X_{TF}$ in the following form
\begin{align}\nonumber
&\bigg[4\pi\bigg({\rho}+\frac{\mathcal{T}_{00}^{(\mathcal{GB})}}{\mathcal{J}^2}\bigg)+\mathcal{X}_{TF}
\bigg]\dot{}+\frac{1}{3}(2\mathcal{X}_{TF}-\mathcal{Y}_{TF}+\mathcal{X}_{\mathcal{T}}+\mathcal{Y}_{\mathcal{T}})(\Theta-\sigma)\mathcal{J}
\\\tag{A10}\label{eq:A10}&+\frac{12\pi \mathcal{J} \mathcal{L}'}{\mathcal{K} \mathcal{L}}\bigg(q-\frac{\mathcal{T}_{01}^{(\mathcal{GB})}}{\mathcal{J} \mathcal{K}}\bigg)-\frac{\dot{\mathcal{L}}}{2\mathcal{L}^3}=0.
\end{align}


\vspace{0.3cm}

\begin{thebibliography}{40}

\bibitem{1}  Buchdahl, H. A.: Mon. Not. R. Astron. Soc.
\textbf{150}(1970)1.

\bibitem{2} Dolgov, A.D. and Kawasaki, M.: Phys. Lett. B \textbf{573}(2003)1;
Capozziello, S., Cardone, V.F. and Troisi, A.: Phys. Rev. D
\textbf{71}(2005)043503.

\bibitem{2a} Sharif, M and Gul, M.Z.: Chin. J. Phys.
\textbf{57}(2019)329; Int. J. Mod. Phys. D \textbf{28}(2019)1950054;
Eur. Phys. J. Plus \textbf{133}(2018)345.

\bibitem{2aa} Adeel, M.: et al.: Mod. Phys. Lett. A \textbf{38}(2023)2350152.

\bibitem{2b} Sharif, M.: et al.: Chin. J. Phys. \textbf{91}(2024)66;
Mod. Phys. Lett. A \textbf{39}(2024)2450140.

\bibitem{2c} Rani, S.: et al.: Int. J. Geom. Methods Mod. Phys.
\textbf{21}(2024)2450033.

\bibitem{2d} Gul, M.Z. et.: Eur. Phys. J. C \textbf{84}(2024)8; ibid 1232.

\bibitem{2e} Maurya, S.K. et al.: Phys. Dark Universe \textbf{46}(2024)101619.

\bibitem{2f} Sharif, M.: et al.:  Chin. J. Phys.
\textbf{91}(2024)66; Chin. J. Phys. \textbf{89}(2024)266.

\bibitem{2g} Heisenberg, L.: Phys. Rep. \textbf{1066}(2024)78.

\bibitem{2h} Sharif, M.: et al.: Eur. Phys. J. C
\textbf{84}(2024)1065; ibid., 1094

\bibitem{2i} Koussour, M. et al.: Phys. Dark Universe
\textbf{46}(2024)101577.

\bibitem{2l} Myrzakulov, Y. et al.: Phys. Dark Universe
\textbf{45}(2024)101545; J. High Energy Astrophys. \textbf{44}
(2024)164.

\bibitem{2m} Sharif, M. and Gul, M.Z..: Ann. Phys.
\textbf{465}(2024)169674; Phys. Dark Universe
\textbf{46}(2024)101606.

\bibitem{2n} Gul, M.Z. et al.: Eur. Phys. J. C
\textbf{84}(2024)802; ibid 775.

\bibitem{2o} Sharif, M. and Gul, M.Z.: Phys. Scr.
\textbf{99}(2024)065036.

\bibitem{2p} Errehymy, A. et al.: Phys. Dark Universe \textbf{46}(2024)101555.

\bibitem{2qq} Nan, G. et al.: Phys. Dark Universe
\textbf{46}(2024)101635.

\bibitem{2qqq} Gul, M.Z. et al.: Fortschr. Phys.
\textbf{72}(2024)2300221.

\bibitem{2r} Koussour, M.: Phys. Dark Universe \textbf{45}(2024)101527.

\bibitem{2rrr} Sharif, M. et al.: Phys. Scr. \textbf{99}(2024)115003.

\bibitem{2ss} Sharif, M. et al.: Mod. Phys. Lett. A
\textbf{39}(2024)2450140.

\bibitem{2sss} Rani, S.: et al.: Phys. Dark Universe \textbf{47}(2025)101754.

\bibitem{2ssss} Sharif, M. et al.: Phys. Dark Universe \textbf{47}(2025)101760.

\bibitem{3} Deruelle, N.: Nucl. Phys. B \textbf{327}(1989)253;
Deruelle, N. and Farina-Busto, L.: Phys. Rev. D
\textbf{41}(1990)3696.

\bibitem{4} Bhawal, B. and Kar, S.: Phys. Rev. D
\textbf{46}(1992)2464; Deruelle, N. and Dolezel, T.: Phys. Rev. D
\textbf{62}(2000)103502.

\bibitem{5} Nojiri, S. and Odintsov, S.D.: Phys. Lett. B \textbf{631}(2005)1.

\bibitem{6} De Felice, A., Hindmarsh, M. and Trodden, M.: J. Cosmol. Astropart. Phys. \textbf {08}(2006)005.

\bibitem{7} Cognola, G. et al.: Phys. Rev. D
\textbf{73}(2006)084007; Amendola, L., Charmousis, C. and Davis,
S.C.: J. Cosmol. Astropart. Phys. \textbf{10}(2007)004; Amendola,
L., Charmousis, C. and Davis, S.C.: Gravit. \textbf{49}(2017)112.

\bibitem{7a} Donmez, O.: Eur. Phys. J. C \textbf{81}(2021)113.

\bibitem{7b} Donmez, O., Dogan, F. and Sahin, T.: Universe \textbf{8}(2022)458.

\bibitem{7c} Babar, R. et al.: Int. J. Mod. Phys. A \textbf{38}(2023) 2350035.

\bibitem{7d} Donmez, O.: Phys. Dark Universe \textbf{1}(2024)101763.

\bibitem{7dd} Gul, M.Z. and Sharif, M.: Chin. J. Phys. \textbf{88}(2024)388.

\bibitem{7e} Donmez, O.: Res. Astron. Astrophys.
\textbf{24}(2024)085001.

\bibitem{7f} Donmez, O.: Mod. Phys. Lett. A \textbf{39}(2024)2450076.

\bibitem{8} Lopez-Ruiz, R., Mancini, H.L. and Calbet, X.: Phys. Lett. A
\textbf{209}(1995)321.

\bibitem{9} Calbet, X. and Lopez-Ruiz, R.: Phys. Rev. E
\textbf{63}(2001)066116; Catalan, R.G., Garay, J. and Lopez-Ruiz, R.
Phys. Rev. E \textbf{66}(2002)011102.

\bibitem{10} Sanudo, J. and Lopez-Ruiz, R.: Phys. Lett. A \textbf{372}(2008)5283; De
Avellar, M.G.B. et al.: Phys. Lett. A \textbf{378}(2014)3481.

\bibitem{11} Herrera, L.: Phys. Rev. D \textbf{97}(2018)044010.

\bibitem{13} Herrera, L., Di Prisco, A. and Ospino, J.:  Phys. Rev. D
\textbf{98}(2018)104059.

\bibitem{14} Herrera, L., Di Prisco, A. and Ospino, J.: Phys. Rev. D
\textbf{99}(2019)044049.

\bibitem{15} Contreras, E., Fuenmayor, E. and Abellan, G.: Eur. Phys. J. C
\textbf{82}(2022)187.

\bibitem{16} Contreras, E. and Stuchlik, Z.: Eur. Phys. J. C \textbf{82}(2022)706.

\bibitem{17} Abbas, G. and Nazar, H.: Eur. Phys. J. C \textbf{78}(2018)957.

\bibitem{18} Nasir, M.M.M. et al.: Eur. Phys. J. C \textbf{85}(2025)159.

\bibitem{21} Herrera, L. and Di Prisco, A.: J. Ospino, Gen. Relativ. Gravit. \textbf{44}(2012)
2645.

\bibitem{22} Houndjo, M.J.S. et al.: Can. J. Phys. \textbf{92}(2014)1528.

\bibitem{23} Sharif, M. and Butt, I.I.: Eur. Phys. J. C
\textbf{78}(2018)850.

\bibitem{25} Nasir, M.M.M., Bhatti, M.Z. and Yousaf, Z.: Int. J. Mod. Phys. D \textbf{32}(2023)2350009.

\bibitem{33} Nojiri, S. and  Odintsov, S.D.: Phys. Lett. B \textbf{631}(2005)1.

\bibitem{33a} Sharif, M. and Naseer, T.:
Chin. J. Phys. \textbf{77}(2022)2655.

\bibitem{33b} Thorne, K.S.: Phys. Rev. \textbf{138}(1965)B251; ibid.
\textbf{139}(1965)B244.

\bibitem{35} Misner, C.W. and  Sharp, D.H.: Phys. Rev. \textbf{136}(1964)B571.

\bibitem{36} Herrera, L. et al.: Phys. Rev. D \textbf{79}(2009)064025.

\bibitem{37} Herrera, L. et al.: Gen. Relativ. Gravit. \textbf{46}(2014)1827.

\bibitem{38} Herrera, L.,  Di Prisco, A. and Ospino, J.: Phys. Rev. D
\textbf{98}(2018)104059.

\bibitem{39} Herrera, L,. Di Prisco, A. and Ospino, J.: Phys. Rev. D
\textbf{99}(2019)044049.

\bibitem{40} Herrera, L,. Di Prisco. and Ospino.: Gen. Relativ. Gravit. \textbf{42}(2010)1585.

\bibitem{41} Herrera, L. et al.: Phys. Rev. D \textbf{69}(2004)084026.

\bibitem{42} Thirukkanesh, S. and Maharaj, S.D.: J. Math. Phys. \textbf{51}(2010)7.

\bibitem{43} Israel, W. and Stewart, J.M.: Phys. Lett. A \textbf{58}(1976)215.

\end{thebibliography}
\end{document}